\pdfoutput=1

\documentclass[aps,prl,10pt,twocolumn,showpacs,superscriptaddress,footnoteinbib]{revtex4}
\usepackage{amsmath}
\usepackage{latexsym}
\usepackage{amssymb}
\usepackage{graphics,epstopdf}

\begin{document}

\title{Fine-grained lower limit of entropic uncertainty in the presence of 
quantum memory}

\author{T. Pramanik}
\thanks{tanu.pram99@bose.res.in}

\author{P. Chowdhury}
\thanks{priyanka@bose.res.in} 

\author{A. S. Majumdar}
\thanks{archan@bose.res.in}
\affiliation{S. N. Bose National Centre for Basic Sciences, Salt Lake, Kolkata 700 098, India}

\date{\today}

\begin{abstract}
 
The limitation on obtaining precise outcomes of measurements 
performed on two non-commuting observables of a particle as set by the 
uncertainty principle in its entropic form, can be reduced in the presence
of quantum memory. We derive a new entropic uncertainty relation based on 
fine-graining, which leads to an ultimate limit on the precision achievable 
in measurements performed on two incompatible observables in the presence of 
quantum memory. We show that our derived uncertainty relation tightens the 
lower bound set by entropic uncertainty for members of the class of two-qubit 
states with maximally mixed marginals, while accounting for the recent 
experimental results using maximally entangled pure states and mixed 
Bell-diagonal states. An implication of our uncertainty relation on the 
security of quantum key generation protocols is pointed out.

\end{abstract}


\maketitle

In the absence of quantum memory, the Heisenberg uncertainty 
principle\cite{HUR} bounds the product of uncertainties, i.e., the spread 
measured by standard deviation, of measurement outcomes for two 
non-commutating observables. The Heisenberg uncertainty principle, for two
 observables $R$ and $S$, is given by
\begin{eqnarray}
\Delta R.\Delta S \geq \frac{1}{2} |\langle [R,S]\rangle|
\label{HUR}
\end{eqnarray}
where $\Delta R$ ($\Delta S$) represents the standard deviation which is a 
measure of uncertainty of the corresponding observable $R (S)$. The possibility
of violating the uncertainty principle using quantum entanglement was one
of the off-shoots of the famous Einstein-Podolsky-Rosen argument\cite{EPR}.
An experiment to demonstrate the violation of the uncertainty principle
was proposed by Popper\cite{popper}, and subsequently realized much later
by Kim and Shih\cite{kim}. Other experiments using entangled states to 
demonstrate the violation of the Heisenberg uncertainty principle have
also been performed\cite{hofmann,bowen,howell}.

There is an increasing appreciation in recent times of the limitations of 
the use of standard 
deviation as a measure of uncertainty\cite{Drawback}. One of the drawbacks of 
the uncertainty relation in terms of standard deviation is that the right hand 
side of the inequality (\ref{HUR}) depends on the state of the quantum system. 
To improve this situation as well as link uncertainty with information 
theoretic concepts, the uncertainty relating to the outcomes of observables 
has been reformulated in terms of Shannon entropy\cite{bbl} instead of
standard deviation. Entropic uncertainty relations for two observables 
was first introduced by Deutsch\cite{EUR1}, following which  
an improved version given by 
\begin{eqnarray}
\mathcal{H}(R)+\mathcal{H}(S) \geq \log_2 \frac{1}{c}
\label{EUR1}
\end{eqnarray}
was first conjectured\cite{EUR2}, and then proved\cite{EUR3}. 
Here $\mathcal{H}(i)$ denotes the Shannon entropy of the probability 
distribution of the measurement outcomes of observable $i$ ($i\in\{R,S\}$) and 
$\frac{1}{c}$ quantifies the complementarity of the observable. For 
non-degenerate observables, $c=\max_{i,j} |\langle a_i|b_j\rangle|^2$, 
where $|a_i\rangle$ and $|b_j\rangle$ are eigenvectors of $R$ and $S$, 
respectively.

In a recent work, Berta et al. \cite{QMemory} have shown that the lower bound 
of entropic uncertainty  (given by Eq.(\ref{EUR1})) can be improved 
in the presence of quantum memory, making use of the quantum information
contained in the correlated state of the particle on which the two observables 
are measured, with the state of another particle. Specifically, 
the sum of uncertainties of two measurement outcomes ($\mathcal{H}(R)+\mathcal{H}(S)$) for measurement of two observables $(R,S)$ on the quantum system (``A", possessed by Alice) can be reduced to $0$ (i.e., there is no uncertainty) if 
that system is maximally entangled with an another system, called quantum 
memory (``B", possessed by Bob). Here, Bob is able to reduce his uncertainty 
about Alice's measurement outcome with the help of communication from Alice 
regarding the choice of her 
measurement performed, but not its outcome. The entropic uncertainty relation 
in the presence of quantum memory\cite{QMemory} is given by
\begin{eqnarray}
\mathcal{S}(R|B)+\mathcal{S}(S|B) \geq \log_2 \frac{1}{c} + \mathcal{S}(A|B)
\label{EUR-QM}
\end{eqnarray}
where $\mathcal{S}(R|B)$ ($\mathcal{S}(S|B)$) is the conditional von Neumann 
entropy of the state given by
$\sum_{j} (|\psi_j\rangle\langle\psi_j|\otimes I)\rho_{AB}(|\psi_j\rangle\langle\psi_j|\otimes I)$,
with $|\psi_j\rangle$ being the eigenstate of observable $R  (S)$, and 
$\mathcal{S}(R|B)$ ($\mathcal{S}(S|B)$) quantifies the uncertainty corresponding to the measurement $R (S)$ on the system ``A" given information stored in the 
system ``B" (i.e., quantum memory).  $\mathcal{S}(A|B)$ quantifies the amount 
of entanglement between the quantum system possessed by Alice and the quantum 
memory possessed by Bob. For example, if Alice and Bob share a maximally 
entangled state, $\mathcal{S}(A|B)=-1$,  and  for a two qubit case  
$\log_2 \frac{1}{c}$ can not larger than $1$, and hence the right hand side of
equation (\ref{EUR-QM}) cannot be greater than $0$ for a maximally entangle state.
It follows that for maximally entangled state Bob's uncertainty of Alice's
measurement outcome reduces to zero when Alice measures the same observable
as Bob does on his quantum memory, and 
communicates with Bob about her measurement choice. 

The effectiveness of 
quantum memory in reducing quantum uncertainty has been demonstrated 
in two recent experiments using respectively, pure\cite{Expt.2}  and
mixed\cite{Expt.1} entangled states. 
For the purpose of experimental verification of inequality (\ref{EUR-QM}),
The entropic uncertainty is recast in the form of the sum of the Shannon
entropies $\mathcal{H}(p^R_d) + \mathcal{H}(p^S_d)$ when Alice and Bob measure the same observables
$R(S)$ on their respective systems and get different outcomes whose 
probabilities are denoted by $p^R_d(p^S_d)$, 
and $\mathcal{H}(p^{R(S)}_d) = -p^{R(S)}log_2p^{R(S)} -(1-p^{R(S)})log_2(1-p^{R(S)})$.  
Making use of Fano's 
inequality\cite{Fano},
it follows that $\mathcal{H}(p^R_d) + \mathcal{H}(p^S_d) \ge \mathcal{S}(R|B)+\mathcal{S}(S|B)$
which using  Eq.(\ref{EUR-QM}) gives \cite{Expt.1}
\begin{eqnarray}
\mathcal{H}(p^R_d) + \mathcal{H}(p^S_d) \ge  \log_2 \frac{1}{c} + \mathcal{S}(A|B)
\label{EUR-QM2}
\end{eqnarray}
The right hand side of the inequality (\ref{EUR-QM2}) can be
determined from the knowledge of the state and the measurement settings.
It was experimentally observed by Li et al. \cite{Expt.1} that the left hand side exceeds the
right hand side for the case of a Bell-diagonal state.
It may be noted that the lower bound
of entropic uncertainty given by the right hand
sides of the relations (\ref{EUR1}) and (\ref{EUR-QM}) contain
the term $1/c$ which depends on the choice of measurement settings.

A further improvement in the manifestation of the uncertainty in measurement
outcomes has been motivated by the realization that entropic
functions provide a rather coarse way of
measuring the uncertainty of a set of measurements,
as they do not distinguish the uncertainty
inherent in obtaining any combination of outcomes
 for different measurements. In the same year of the work by
Berta et al. \cite{QMemory}, a new form of the uncertainty relation, {\it viz.},
{\it fine grained uncertainty relation}, was proposed by Oppenheim and Wehner \cite{FUR1}.  
 In particular, they considered
a game according to which Alice and Bob both receive binary questions, 
i.e., projective spin measurements along two different directions at each side.
The winning probability is given by the
relation \cite{FUR1}
\begin{eqnarray}
P^{game}(\mathcal{T}_A,\mathcal{T}_B,\rho_{AB})= \\
\displaystyle\sum_{s,t} p(t_A,t_B) \displaystyle\sum_{a,b} V(a,b|t_A,t_B) \langle (A_{t_A}^a\otimes B_{t_B}^b) \rangle_{\rho_{AB}}
&\leq & P^{game}_{max} \nonumber 
\label{FUR1}
\end{eqnarray}
where $\rho_{AB}$ is a  bipartite state shared by  Alice and Bob, and $\mathcal{T}_A$ and $\mathcal{T}_B$ represent the set of measurement settings $\{t_A\}$ and $\{t_B\}$ chosen by Alice and Bob, respectively, with probability $p(t_A,t_B)$. 
Alice's (Bob's) question and answer are $t_A (t_B)$ and $a (b)$, 
respectively,  with 
$A_{t_A}^a=\frac{[I+(-1)^a A_{t_A}]}{2}$ ($B_{t_B}^b=\frac{[I+(-1)^b B_{t_B}]}{2}$) 
being a measurement of the observable $A_{t_A}$ ($B_{t_B}$). Here 
$V(a,b|t_A,t_B)$ is some function determining the winning condition of the 
game. The necessary condition for
fine-graining is to consider a particular outcome, or particular choice of
the winning condition in a game. The winning condition is the essence of 
fine-graining, and as shown in Ref.\cite{FUR1}, every game gives rise to an
uncertainty relation, and vice-versa.

The winning
condition corresponding to  a special class of nonlocal retrieval games 
(CHSH game \cite{FUR1}) for which there exist only 
one winning answer for one of the two parties, is given by $V(a,b|t_A,t_B)=1$, iff $a\oplus b=t_A.t_B$, and 
 $0$ otherwise. $P^{game}_{max}$ is the maximum winning probability of the game, 
maximized over the set of projective spin measurement settings $\{t_A\}$ ($\in$ $\mathcal{T}_A$) by Alice, the set of projective spin measurement settings $\{t_B\}$ ($\in$ $\mathcal{T}_B$) by Bob, 
i.e.,
$P^{game}_{max}=\max_{\mathcal{T}_A,\mathcal{T}_B,\rho_{AB}} P^{game}(\mathcal{T}_A,\mathcal{T}_B,\rho_{AB})$.
Using the maximum winning probability it is possible to 
discriminate between classical theory, quantum theory and no-signaling 
theory with the
help of the degree of nonlocality \cite{FUR1}.
A generalization 
to the case of tripartite systems has also been proposed \cite{FUR2}.

In the present work we derive a new form of the uncertainty relation 
in the presence of quantum memory, in which the lower bound of entropic 
uncertainty corresponding to the
measurement of two observables is determined by fine-graining of the
possible measurement outcomes, and is thus independent of the specific
choice of measurement settings. We find the finer or optimized lower 
bound of entropic uncertainty, which represents the ultimate limit to
which uncertainty of outcomes of two non-commuting observables can be
reduced by performing any set of measurements  in the presence of quantum 
memory. The new uncertainty relation derived by us is able to account for
the two experimental results obtained for the case of maximally entangled
states\cite{Expt.2} and mixed Bell-diagonal states\cite{Expt.1}.
More interestingly, we show that when the quantum correlations are made
using the class of two-qubit states with maximally mixed marginals, the
fundamental limit set by our uncertainty relation prohibits the attainment
of the lower bound of entropic uncertainty\cite{QMemory} as defined by
the right hand side of  equation (\ref{EUR-QM}). We further discuss the 
ramification of our uncertainty relation on an application for key extraction
in quantum key generation\cite{QMemory,devetak,renes}. 

We consider
a quantum game played by Alice and Bob for which the winning 
probability is given by the fine-grained uncertainty relation \cite{FUR1}.  
In this game, 
Alice and Bob share a two-qubit state $\rho_{AB}$ which is prepared by Alice,
after which she sends one of the qubits to Bob. Bob's qubit represents the 
quantum memory, and mimicking the scenario of references \cite{QMemory,Expt.1,Expt.2} we look at Bob's uncertainty of 
the outcome of Alice's measurement of one of two incompatible observables (say, $R$ and $S$), when Alice helps Bob  by communicating her measurement choice of 
a suitable spin observable on her system. In this game Alice and Bob are 
driven by the requirement of minimizing the value of the quantity 
$\mathcal{H}(p^R_d) + \mathcal{H}(p^S_d)$ which forms the left hand side of the entropic 
uncertainty relation (\ref{EUR-QM2}). The minimization is over all 
incompatible measurement settings
such that $R \neq S$,
i.e., 
\begin{eqnarray}
\mathcal{H}(p^R_d) + \mathcal{H}(p^S_d) \ge \min_{R, S\neq R}[\mathcal{H}(p^R_d) + \mathcal{H}(p^S_d)]
\label{step1}
\end{eqnarray} 
To find
the minimum value, the choice of one variable, e.g., $R$, may be fixed,
without the loss of generality to be, say $\sigma_z$ (spin measurement 
along the $z$-direction), and then the minimization be performed over the
other variable $S$. Hence, Eq.(\ref{step1}) can be rewritten in the form
\begin{eqnarray}
\mathcal{H}(p^R_d) + \mathcal{H}(p^S_d) \ge \mathcal{H}(p^{\sigma_z}_d) +
\min_{S\neq \sigma_z}[ \mathcal{H}(p^S_d)]
\label{step2}
\end{eqnarray} 
The uncertainty defined by the entropy $ \mathcal{H}(p^S_d)$ is minimum when
the certainty of the required outcome is maximum, corresponding to an
infimum value for the probability $p^S_d$.

Now, in order to obtain the infimum value of $p^S_d$, we use the fine-grained
uncertainty relation \cite{FUR1} in a form relevant to the present situation.
In the language of the above bipartite game, Alice and Bob measure the same
observables ($\sigma_z$ and $S$) on their respective systems, and win the
game if their measurement outcomes, either $0$ or $1$ ($a$ for Alice, and $b$ for Bob) are
correlated in the form $a \oplus b = 1$, i.e., they get different outcomes. 
Therefore, the fine-grained uncertainty relation (5) in the context
of this particular game considered here ($s=t$ in Eq.(5)), will now 
be determined by a new
winning condition as given below where the
infimum value of the winning probability (corresponding to minimum uncertainty)
is given by 
\begin{eqnarray}
p_{\inf}^{S}= \inf_{S(\ne \sigma_z)}\displaystyle\sum_{a,b} V(a,b) Tr[(A_S^a\otimes B_S^b).\rho_{AB}], 
\label{inf-S}
\end{eqnarray}
with the winning condition $V(a,b)$ given by
\begin{eqnarray}
V(a,b) &=& 1 \textit{~~~~~ iff $a\oplus b=1$} \nonumber \\
       &=& 0 \textit{~~~~~ otherwise}.
\end{eqnarray}
with $A_S^a$ being a projector for observable $S$ with outcome `$a$', given by
$S^{\alpha}=\frac{I+(-1)^{\alpha} \vec{n}_{S}.\vec{\sigma}}{2}$ (and similarly for
$B_S^b$),
where $\vec{n}_{S}(\equiv \{\sin(\theta_{S}) \cos(\phi_{S}), \sin(\theta_{S}) \sin(\phi_{S}),$ $\cos(\theta_{S}) \} )$; $\vec{\sigma}\equiv \{\sigma_x,\sigma_y,\sigma_z\}$ are the Pauli matrices; $\alpha$ takes the value either $0$ (for spin 
up projector) or $1$ (for spin down projector). Note here that the above
winning condition proposed by us is different from the winning condition
used in ref. \cite{FUR1} for the purpose of capturing the nonlocality of
bipartite systems. The essence of fine-graining is to consider a particular 
outcome, or particular choice of the winning condition in a game. We have
adapted the fine-grained uncertainty
relation making it directly applicable to the experimental situation of quantum
memory, by introducing a new winning condition modelling 
the experiments  \cite{Expt.1}.

The minimum value of uncertainty thus obtained by minimizing over all
measurements is now substituted in the second term of the right hand side of
Eq.(\ref{step2})
from which the expression for the final form of our uncertainty relation
\begin{eqnarray}
\mathcal{H}(p^R_d)+\mathcal{H}(p^S_d) \geq  \mathcal{H}(p^{\sigma_z}_d) +
\mathcal{H}(p^S_{inf})
\label{FURQM1}
\end{eqnarray}
follows giving the optimal lower bound of entropic uncertainty. The
value of $p^S_{inf}$ is calculated for the given quantum state $\rho_{AB}$
using the expression (\ref{inf-S}). 
As a result, the lower bound of the entropic uncertainy in the presence
of quantum correlations is now determined 
by the minimum entropy corresponding to the infimum winning probability of 
the above
game, replacing the earlier lower bound given by the right hand side of 
 Eq.(\ref{EUR-QM2}) \cite{QMemory,Expt.1}. Note that the inequality 
(\ref{FURQM1}) can
be derived for any choice of $R$ other than $\sigma_z$ as well. Our proposed 
uncertainty relation is independent of the choice of measurement settings as it 
optimizes the reduction of uncertainty
quantified by the conditional Shannon entropy over all possible observables.
Given a bipartite state possessing quantum correlations, inequality 
(\ref{FURQM1}) provides the fundamental limit to which uncertainty in the
measurement outcomes of any two incompatible variables can be reduced.

In the following analysis, we illustrate the efficacy of our uncertainty
relation (\ref{FURQM1}) with some examples. We use 
Eq.(\ref{inf-S}) to first calculate the value of  $p^S_{inf}$ 
(the optimization over all 
spin projectors is performed using {\it Mathematica}) and the corresponding
measurement setting $S$, and then use it to find the minimum value or
lower bound of uncertainty defined by the right hand side of the equation
(\ref{FURQM1}) for the examples of the different states representing quantum
memory discussed here, {\it viz.}, maximally entangled state, Bell-diagonal 
state, an example from the class of two-qubit states with maximally mixed
marginals, and the Werner state. We further describe how our derived
uncertainty relation affects an
important application to quantum key distribution\cite{ekert} modifying 
the earlier
bounds\cite{devetak} on the amount of key per state that Alice and 
Bob are able to
extract\cite{QMemory,renes}

First, we consider that Alice and Bob share a 
maximally entangled state. For any maximally entangled state the outputs are 
strongly correlated when both Alice and Bob measure the same observable on 
their respective systems. When Alice communicates about 
her measurement setting,
 Bob knows with certainty about Alice's outcome by measuring the same 
observable on his system. The lower bound of  entropic uncertainty should
thus reduce to zero in this case, as observed earlier\cite{QMemory,Expt.2}.
Using our uncertainty relation (\ref{FURQM1}), it indeed follows that
$\mathcal{H}(p^{\sigma_z}_d) + \mathcal{H}(p^S_{inf}) =0$, 
a result which holds for any choice of the observable $S$, as long as Alice
and Bob measure the same observables. 
Now, for an observable parametrized by
$S = \hat{n}_S.\vec{\sigma}$,
where $\hat{n}_S (\equiv \{\sin(\theta_{S}) \cos(\phi_{S}), \sin(\theta_{S}) \sin(\phi_{S}),$ $\cos(\theta_{S}) \})$,
note that the right hand side of the entropic uncertainty relation 
(\ref{EUR-QM2}) is given by $\log_2\frac{1}{c} + \mathcal{S}(A|B)=-1 + \log_2[\frac{1}{\max[\cos(\frac{\theta_S}{2})^2,\cos(\frac{\theta_S}{2})^2]}]$
which varies from $0$ to $-1$ depending upon the measurement settings 
$\{\theta_S,\phi_S\}$, and hence the lower bound of the uncertainty relation given by
Berta et al. \cite{QMemory} for a maximally 
entangled state is $0$ only when the observables $R$ and $S$ are complementary 
to each other, i.e., $c=\max_{i,j} |\langle a_i|b_j\rangle|^2=\frac{1}{2}$,
as observed experimentally with horizontal and vertical polarized photons 
by Prevedel et al.\cite{Expt.2}.

Next, let us consider the Bell-diagonal state (used in the experiment by Li et al. \cite{Expt.1}) given by
$\rho_m=p |\phi^{+}\rangle\langle\phi^{+}|+(1-p) |\psi^{-}\rangle\langle\psi^{-}|$,
where, $|\phi^{+}\rangle=\frac{1}{\sqrt{2}} (|00\rangle+|11\rangle)$, and $p$  
lies between $0$ and $1$. The lower bound of entropic uncertainty  in the 
presence of quantum memory (given by inequality (\ref{EUR-QM2})) for the state 
$\rho_m$ with the choice of observables $S=\sigma_x$ and $R= \sigma_z$ 
($\theta_S = \pi/2, \phi_S = 0$) 
is given by $\mathcal{H}(p) \equiv \mathcal{H}(1-p)$. It can be
verified that the lower bound of our uncertainty relation (\ref{FURQM1}) 
is obtained for
the same variables with the corresponding probabilities given by 
$p^{\sigma_z} = 1-p$ and $p^S_{inf} \equiv p^{\sigma_y} = 1$, leading to saturation
of the bound
$\mathcal{H}(p^R_d)+\mathcal{H}(p^S_d) \geq \mathcal{H}(1-p)$
for the Bell-diagonal state. Note here that the choice $R=\sigma_z$ $S=\sigma_x$
(as taken by  Li et al. \cite{Expt.1}) is unable to minimize the left hand 
side of the above expression,  and thus we account for  the fact that
their experimental result (left hand side of the inequality (4)) is
obtained to lie above the lower bound.

\begin{figure}[!ht]
\resizebox{7cm}{6cm}{\includegraphics{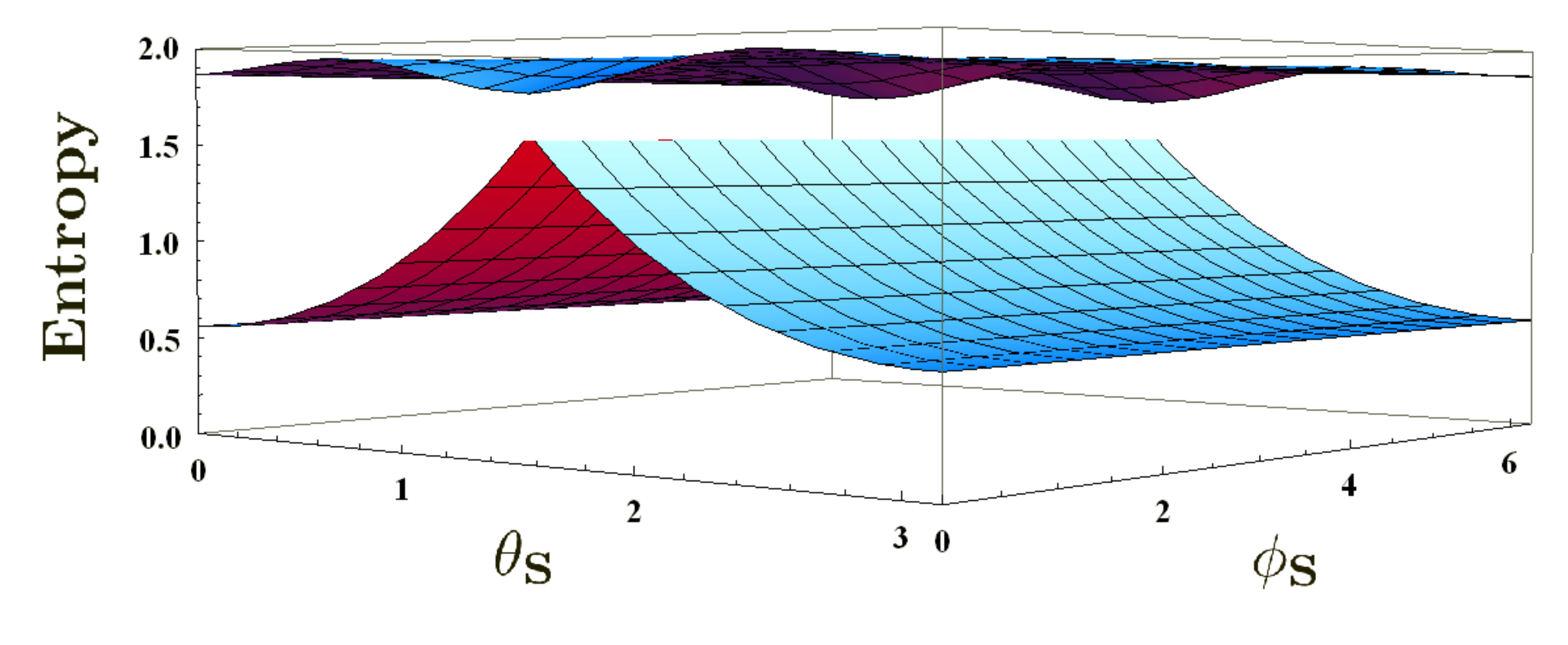}}
\caption{\footnotesize Coloronline. The lower bound of entropic uncertainty corresponding to 
measurements on a two-qubit state with maximally mixed marginals
in the presence of quantum memory.
(i) the upper plot $\mathcal{H}(p^{\sigma_z}_d) + \mathcal{H}(p^S_{inf})$ as 
predicted by using our 
uncertainty relation (\ref{FURQM1}) derived here, and (ii) the lower plot 
$\log_2\frac{1}{c} + \mathcal{S}(A|B)$ as predicted by the analysis of
Berta et al.\cite{QMemory} given by (\ref{EUR-QM}). The region between
the two curves is inaccessible in actual measurements according to our
results, since the optimal lower bound of entropic uncertainty is determined
by fine-graining.}
\end{figure}

Finally, we consider the general class of two-qubit states with maximally
mixed marginals, given by
$\rho_{MM}=\frac{1}{4} (I_{4\times 4}+\displaystyle\sum_{i=1}^3 c_i \sigma_i\otimes \sigma_i)$,
where the $c_i$ are real constants satisfying the constraints 
$0\leq \frac{1-c_1-c_2-c_3}{4}\leq 1, 
0\leq \frac{1+c_1-c_2-c_3}{4}\leq 1,
0\leq \frac{1-c_1+c_2-c_3}{4}\leq 1, 
0\leq \frac{1-c_1-c_2+c_3}{4}\leq 1$,
such that the state $\rho_{MM}$ is physical.  We obtain the lower bound of the 
inequality 
(\ref{FURQM1}), and the corresponding setting $S$ which is then used to 
compare the  
minimum of entropy thus obtained with the lower bound of 
the inequality (\ref{EUR-QM2}). For a wide range of choices of the state 
parameters $c_i$ we find that the fundamental limit set by the inequality
(\ref{FURQM1}) as obtained through fine-graining exceeds the lower bound
applying the right hand side of  equation (\ref{EUR-QM2}). A typical example, using
the values $c_1 = 0.5, c_2 = -0.2, c_3 = -0.3$, is illustrated in Fig.1.
Note that the minimum value of $\mathcal{H}(p^R_d)+\mathcal{H}(p^S_d)$ occurs
in this case for $\theta_S = \pi/2, \phi_S=0$, yeilding the lower bound
of (\ref{FURQM1}), 
$\mathcal{H}(p^{\sigma_z}_d) + \mathcal{H}(p^S_{inf}) \approx 1.745$, while for
the same observables, one obtains the
right hand side of (\ref{EUR-QM2}) as $
 \log_2 \frac{1}{c} + \mathcal{S}(A|B) \approx 1.558$. It is seen from Fig.1
that when this specific state is used as quantum memory, the lower bound
of entropic uncertainty as predicted by the analysis of Berta
et al. \cite{QMemory} is never achievable in an actual experiment
using any choice of the measurement settings $\{\theta_S, \phi_S\}$. As
a further illustration of these results, one may also consider the Werner 
state $\rho_W=\frac{1-p}{4} I\otimes I+ p |\psi^{-}\rangle\langle\psi^{-}|$. 
Here fine-graining leads to the lower bound $2\mathcal{H}(\frac{1+p}{2})$
which always exceeds the right hand side of the inequality (\ref{EUR-QM2}) 
($- 3 \frac{1-p}{4} \log_2 \frac{1-p}{4} - \frac{1+3p}{4} \log_2 \frac{1+3p}{4}$), except for $p=0$ (maximally mixed state leading to
maximum and equal uncertainty using both the approaches), and for $p=1$
(maximally entangled state leading to vanishing uncertainty in both approaches). 

The uncertainty principle in its entropic form could be used
for verifying the security of key distribution protocols. It was derived
by Devetak and Winter \cite{devetak} that the amount of key $K$ that Alice and Bob are able to
extract per state  should always exceed the quantity $\mathcal{S}(R|E)
- \mathcal{S}(R|B)$, where the quantum state  $\rho_{ABE}$ is shared between 
Alice, Bob and the evesdropper Eve ($E$)\cite{ekert}. Extending this idea
by incorporating the effect of shared quantum correlation between Alice and
Bob, Berta et al.\cite{QMemory} reformulated their result (\ref{EUR-QM})
in the form of $\mathcal{S}(R|E) + \mathcal{S}(R|B) \ge \log_2 \frac{1}{c}$
conjectured earlier\cite{renes}, enabling them to derive a new lower bound
on the key extraction rate, given by $K \ge \log_2 \frac{1}{c} - \mathcal{S}(R|B)- \mathcal{S}(S|B)$. Now, using our uncertainty relation (\ref{FURQM1}), it
is possible to obtain a tighter lower bound, given by $K \ge \log_2 \frac{1}{c} -
\min_{R,S}[\mathcal{H}(p^R_d) + \mathcal{H}(p^S_d)]$ which reduces to the
form $K \ge \log_2 \frac{1}{c} - \mathcal{H}(p^{\sigma_z}_d) + \mathcal{H}(p^S_{inf})$ when Alice and Bob retain  data whenever they make the same choice
of measurement on their respective sides. Note that the bound derived here is
upper-bounded by the result of Berta et al.\cite{QMemory}.
The implication  is that the
saturation of the bound derived earlier\cite{QMemory} is not possible
for all states, and the bound derived here represents the optimal lower limit
of key extraction valid for any shared correlation, and for all measurement 
settings used by Alice and Bob.

To summarize, in the present work we give the optimized lower bound of 
entropic uncertainty  in the presence of quantum memory \cite{QMemory}
with the help of the
fine-grained uncertainty principle \cite{FUR1}, thus
providing a new manifestation of observer dependence \cite{winter} of the 
fundamental limitation. Since entropy (or uncertainty) is directly related to
probability, the purpose of minimizing probability as we have done while
implementing the fine-graining, is essential to minimize
uncertainty.  So, we are able in this way to obtain the optimal lower bound
of entropic uncertainty in presence of quantum memory.
In  measurements and communication involving
two parties, the lower bound of entropic uncertainty cannot fall below
the bound derived here, as we illustrate with several examples.  Our 
uncertainty relation  is independent of measurement settings, 
providing an operationally relevant fundamental limitation on the precision
of outcomes for measurement of two incompatible observables in the presence
of quantum memory. Implications on  information processing exist, as 
discussed for the issue of privacy of quantum key generation.

{\it Acknowledgements:}   TP and PC thank UGC, India for financial support. 
ASM acknowledges support from the DST project no. SR/S2/PU-16/2007.

\end{document}